# On the discovery of the classical equations for spin motion in electromagnetic field

V. Hushwater [a)]
70 St. Botolph St, apt. 401, Boston, MA 02116

The letter addresses misattribution of the discovery of the classical equation for spin motion in electromagnetic field, known as the BMT equation, named after Bargmann, Michel, and Telegdi. I argue that J. (or Ya.) Frenkel, along with L. H. Thomas, should be considered as a co-discoverer of this equation. He first derived it, in another form, in 1926.

In a paper[1] J. D. Jackson claims that L. H. Thomas is "the true discoverer and expositor" of the equation that describes the relativistic behavior of the spin polarization of a charged particle with magnetic moment moving in a fixed electromagnetic field. This equation is known as the BMT equation, named after Bargmann, Michel, and Telegdi, who derived it in 1959.[2] But, as correctly argued by Jackson, such an equation was derived and discussed by Thomas in a paper published in January 1927.[3]

However, there is an equivalent way of describing spin motion in electromagnetic field in a covariant manner; namely, by using an anti-symmetric tensor as the relativistic generalization of the intrinsic angular momentum observed in the rest-frame of the particle.[2,4] It was J. Frenkel who first proposed such a description and published the derivation of the corresponding equation in May 1926.[5] (Frenkel became interested in this problem after Pauli showed him a short letter by Thomas[6] on the "Thomas factor" (1/2) before its publication.) This was the first published equation on relativistic motion of spin in an electromagnetic field, and that is why BMT cited Frenkel's paper in Z. Phys[5] as a predecessor. Thomas also discussed such a description and a corresponding equation in his 1927 paper mentioned above, which has led some authors to suggest[4] calling this equation the Frenkel-Thomas equation.

But Thomas published this paper seven months after Frenkel's papers, and likely knew about them in the process of his writing: At the end of his paper[3] Thomas wrote, "*Note added later*. Since the above was written Frenkel has published a paper…" However, for this to be true Thomas would have had to finish his paper before May 1926 and wait about half a year before submitting it.

Moreover, in Thomas' paper he refers to and discusses a paper[7] of Heisenberg and Jordan that follows Frenkel's paper in the same 1926 issue of Z. Phys. He discusses it in a body of the paper, not in the "*Note added later*." It most likely means that he saw the issue of Z. Phys., where the papers of Frenkel, and of Heisenberg and Jordan were published, when he was working on his 1927 paper. I don't believe that for some reasons Thomas studied the paper of Heisenberg and Jordan but did not notice the paper of Frenkel that precedes it! Of course there is a possibility that Thomas learned a paper of Heisenberg and Jordan before it was published, but he does not say this explicitly.

Therefore, it looks as though Thomas, being upset by the invasion of Frenkel in his area of research, did not give him enough credit. One can see this also in the fact that Thomas's reference to the Frenkel's paper does not show a year of its publication, while references to all other papers do.

I think that J. Frenkel should be considered along with L. H. Thomas as a co-discoverer of the classical equations for spin motion in electromagnetic field. Given that Frenkel was a Russian physicist, this probably explains why Russian review papers[8,9] on the history of the classical spin theory do not mention Thomas' second paper.

**Addendum**
This submission is a slightly expanded version of my letter published in Am. J. Phys. **82**, 6 (2014). In his response [ibid, p. 7] Jackson addressed some of my arguments. However he did not even mention my observation in the third-to-last paragraph above.